\documentclass[12pt,a4paper]{article}

\usepackage[utf8]{inputenc}
\usepackage{amsmath, amsfonts, amssymb, mathtools, mathrsfs, bbm}

\usepackage{graphicx}
\usepackage{float}
\usepackage{booktabs}
\usepackage{adjustbox}

\usepackage[table,xcdraw]{xcolor}
\definecolor{lightred}{rgb}{1,0.5,0.5}
\definecolor{lightorange}{RGB}{255,178,102}
\definecolor{mediumlightblue}{RGB}{100,170,255}
\definecolor{mediumlightred}{RGB}{230,80,80}
\definecolor{linkequation}{RGB}{0,0,180}

\usepackage[a4paper, width = 170mm, top = 30mm, bottom = 25mm, 
bindingoffset = 0mm]{geometry}

\usepackage{cite}
\usepackage{tikz}
\usepackage{fontawesome}

\usepackage{url}
\usepackage[
colorlinks=true,
linkcolor=blue,
citecolor=red,
urlcolor=blue
]{hyperref}

\newcommand{\beq}{\begin{equation}}
	\newcommand{\eeq}{\end{equation}}

\newcommand{\plainfootnote}[1]{%
	\begingroup
	\renewcommand{\thefootnote}{}
	\phantomsection
	\footnotetext{#1}
	\addtocounter{footnote}{-1}
	\endgroup
}

\makeatletter
\let\SavedEqref\eqref
\renewcommand*{\eqref}[1]{%
	\begingroup
	\hypersetup{linkcolor=linkequation}
	\SavedEqref{#1}%
	\endgroup
}
\makeatother

\begin{document}
	
\begin{center}
	{\large\bf
		L\'{e}vy-index control of spectral singularities and coherent perfect absorption in non-Hermitian space-fractional quantum mechanics
	}
	
	\vspace{0.7cm}
	
	{\small\sf
		Vibhav Narayan Singh \textsuperscript{\textcolor{red}{1}},
		Mohammad Umar \textsuperscript{\textcolor{blue}{2}},
		Mohammad Hasan \textsuperscript{\textcolor{orange}{3}},
		Bhabani Prasad Mandal \textsuperscript{\textcolor{gray}{4}}
	}

\bigskip

\plainfootnote{\textcolor{mediumlightred}{\faEnvelope}\;
\href{mailto:vibhav.ecc123@gmail.com}{vibhav.ecc123@gmail.com},
\href{mailto:vibhav.narayan@iilm.edu}{vibhav.narayan@iilm.edu}}

\plainfootnote{\textcolor{mediumlightblue}{\faEnvelope}\;
\href{mailto:aliphysics110@gmail.com}{aliphysics110@gmail.com},
\href{mailto:mohd.umar@opc.iitd.ac.in}{mohd.umar@opc.iitd.ac.in}}

\plainfootnote{\textcolor{lightorange}{\faEnvelope}\;
\href{mailto:mohammadhasan786@gmail.com}{mohammadhasan786@gmail.com},
\href{mailto:mhasan@isro.gov.in}{mhasan@isro.gov.in}}

\plainfootnote{\textcolor{gray}{\faEnvelope}\;
\href{mailto:bhabani.mandal@gmail.com}{bhabani.mandal@gmail.com},
\href{mailto:bhabani@bhu.ac.in}{bhabani@bhu.ac.in}}

\vspace{0.4cm}

{\em
	\textsuperscript{\textcolor{red}{1}}School of Basic \& Applied Sciences, IILM University, Greater Noida 201308, India\\
	\vspace{0.3em}
	\textsuperscript{\textcolor{blue}{2}}Optics and Photonics Centre, Indian Institute of Technology Delhi, New Delhi 110016, India\\
	\vspace{0.3em}
	\textsuperscript{\textcolor{orange}{3}}Indian Space Research Organisation, Bangalore 560094, India\\
	\vspace{0.3em}
	\textsuperscript{\textcolor{gray}{4}}Department of Physics, Banaras Hindu University, Varanasi 221005, India
}

\vspace{0.8cm}

\noindent{\bf Abstract}
\end{center}

\noindent
We investigate the scattering features of a non-Hermitian rectangular potential within the framework of space-fractional quantum mechanics. Using the Riesz fractional derivative, we analytically derive locus equations for spectral singularities (SSs) and their time-reversed counterparts, coherent perfect absorption (CPA), in a dimensionless complex-potential parameter space. This geometric locus formulation provides a transparent representation of the SS and CPA conditions and enables direct visualization of how fractional quantum dynamics modifies non-Hermitian scattering. We show that reducing the L\'{e}vy index $\alpha$, which enhances nonlocal transport associated with L\'{e}vy-flight dynamics, systematically lowers the gain-loss strength required for the emergence of SSs and CPAs, while increasing the mode index further suppresses this threshold. In addition, for fixed potential parameters, we demonstrate that decreasing $\alpha$ induces a blue shift of the SS energy, in direct agreement with earlier studies. From this perspective, the L\'{e}vy index $\alpha$ emerges as a tunable control knob for SS-CPA settings in fractional non-Hermitian quantum systems. Beyond its quantum-mechanical setting, this study may find applications in fractional waveguides and metamaterials governed by fractional wave equations. This work also bridges the gap between non-Hermitian quantum mechanics and space-fractional quantum mechanics.

\newpage

%==================== Introduction ========================
\section{Introduction}
Spectral singularities (SSs) \cite{mostafazadeh2009spectral, mostafazadeh2011spectral, mostafazadeh2012optical, mostafazadeh2009resonance, longhi2010spectral, samsonov2011scattering}, which are not permissible for Hermitian systems, represent a defining feature for non-Hermitian systems with a continuous spectrum. This is one of the most intriguing findings in non-Hermitian quantum mechanics (NHQM). The SS was first discovered by Naimark \cite{naimark1960investigation} and later, Schwartz coined the term \textit{spectral singularity} while studying them in a specific class of abstract linear operators \cite{schwartz1960some}. SSs correspond to points in the continuous spectra of open quantum systems involving complex potentials, where a continuum wave solution to the Schr\"{o}dinger equation becomes singular. At these points, the system exhibits infinite reflection and transmission coefficients. The physical manifestation of SS has been extensively investigated within the framework of wave scattering phenomena involving complex barrier potentials. Recently, higher-order spectral SSs have been introduced, unifying exceptional points (EPs) and SS, and demonstrating higher-order lasing \cite{xu2023high}.\\
\indent
Further, another significant feature of NHQM is CPA \cite{chong2010coherent, longhi2010backward, wan2011time, hasan2014critical, noh2012perfect, longhi2010pt, zhang2012controlling, gmachl2010suckers}, the time-reversed partner of SS. CPA is the complete extinction of incoming radiation by a complex potential embedded in a physical system supporting wave propagation. The phenomenon is based on the destructive interference of transmitted and reflected waves. The concept was introduced in \cite{chong2010coherent} and first observed experimentally in \cite{wan2011time} for light interacting with absorbing scatterers. SS and CPA are prominent features of NHQM, along with other important aspects such as invisibility \cite{jin2016reciprocal, mostafazadeh2013invisibility}, reciprocity \cite{deak2012reciprocity}, EPs \cite{heiss2004exceptional, heiss2012physics}, and critical coupling (CC) \cite{cai2000observation, tischler2006critically, Ghatak_2015}.\\
\indent
NHQM is an extension of quantum mechanics (QM), and soon after this extension, a new generalization of QM, space fractional quantum mechanics (SFQM), based on its path integral (PI) formalism, was introduced. In the PI formulation of QM \cite{hibbs1965quantum} \cite{zinn2010path}, PIs are computed over Brownian paths, which represent random processes governed by a Gaussian probability distribution and lead to the Schr\"{o}dinger equation. However, the Brownian process is a specific instance within a broader category of random processes known as L\'{e}vy $\alpha$-stable processes. These processes are non-Gaussian and are characterized by the fractional L\'{e}vy index $\alpha$, where $0 < \alpha \leq 2$. When $\alpha = 2$, the Levy process reduces to the Brownian process, indicating that L\'{e}vy paths are essentially Brownian paths in this case. Nick Laskin extended the PI approach in QM by considering path integrals over L\'{e}vy paths \cite{laskin2000fractional,laskin20001fractional,laskin2002fractional}. This modification led to the fractional Schr\"{o}inger equation, establishing the foundation of a new branch of QM known as SFQM. Subsequently, Naber \cite{naber2004time} developed the time-fractional Schr\"{o}dinger equation (TFSE), and Wang and Xu \cite{wang2007generalized} further generalized it to a space-time fractional Schr\"{o}dinger equation.\\
\indent
SFQM has drawn significant attention from researchers and has emerged as a major focus of extensive study \cite{dong2007some, el2019some, kirichenko2018confinement, medina2019nonadiabatic, zhang2015propagation, ghalandari2019wave, yao2018solitons, xiao2018surface, zhang2017resonant, wang2019elliptic, zhang2017unveiling, hasan2018new, khalili2021fractal, baleanu2009solving, golmankhaneh2024fractal}. SFQM is based on the principles of fractional calculus, also known as $F^{\alpha}$-calculus \cite{golmankhaneh2023fractal}, where the derivative is of fractional order. The fractional approach extends beyond QM and has applications in various fields such as astrophysics and cosmology \cite{rasouli2024fractional, rasouli2022inflation, jalalzadeh2022sitter, el2013fractional, el2017wormholes, micolta2023revisiting, landim2021fractional, jalalzadeh2021prospecting, moniz2020fractional, trivedi2024fractional, bidlan2025reconstructing}, quantum field theory \cite{herrmann2008gauge, tarasov2014fractional, calcagni2021quantum}, plasma physics \cite{faridi2021fractional, abdelwahed2022physical} and optics \cite{zhang2016diffraction}. Recently, the experimental realization of the fractional Schr\"{o}dinger equation in the temporal domain has been successfully demonstrated in optical systems \cite{liu2023experimental}. In this work, phase shifts emulating the effect of fractional group-velocity dispersion (FGVD) were generated using a computer-generated hologram. Quantum tunneling through locally periodic delta and rectangular potentials within the framework of SFQM has been investigated in \cite{tare2014transmission}. Furthermore, quantum tunneling in fractal systems has been investigated in Refs. \cite{umar2023quantum,narayan2023tunneling,singh2025quantum,umar2025tunneling} and tunneling through fractal potentials within the fractional framework has also been explored \cite{singh2023quantum}. Tunneling time through both a rectangular potential barrier \cite{hasan2018tunneling} and a periodic rectangular potential barrier \cite{hasan2020tunneling} has also been investigated within SFQM. In this framework, the absence of the Hartman effect is observed.\\
\indent
Despite significant progress in both NHQM and SFQM, research that explores the intersection of these two domains remains relatively scarce. We have coined this interplay as non-Hermitian space fractional quantum mechanics (NHSFQM) \cite{hasan2018new, umar2025tunneling}. Earlier, we investigated the behavior of SSs and CPA for the complex delta potential and complex rectangular potential within the framework of SFQM, where blue shifts of SS energy were observed with decreasing the fractional order $\alpha$ \cite{hasan2018new}. Recently, we examined tunneling time through a complex potential $V_{r} - iV_{i}$ in SFQM, and observed the potential manifestation of the Hartmann effect for a specific combination of $V_{i}$ and $\alpha$ \cite{umar2025tunneling}. These studies represents explorations of non-Hermitian phenomena within the SFQM framework.\\
\indent
In this work, we investigate the presence and evolution of SSs and their time-reversed counterpart, CPA, within the framework of NHSFQM. Starting from a complex scattering rectangular potential $V_{r}+iV_{i}$, and using the conditions for the occurence for SS and CPA, we represented them geometrically as curves in a dimensionless parameter space defined by the ratios of the real and imaginary parts of the potential to the energy. This energy is actually the energy at which the SS and CPA occur. This formulation enables a transparent visualization of how fractional quantum dynamics, governed by the fractional parameter $\alpha$, reshape non-Hermitian scattering behavior. We showed that reducing $\alpha$, which enhances nonlocal transport, lowers the gain-loss strength required for the formation of SS and CPA. Importantly, we demonstrated that the decrease in $\alpha$ induces a blue shift of the SS energy for a fixed potential parameter $V_{r}$ and $V_{i}$, in direct agreement with the earlier study \cite{hasan2018new}. Altogether, our results establish the L\'{e}vy index $\alpha$ as a tunable nonlocal control knob for SS-CPA physics. Beyond its quantum-mechanical setting, the present study will find natural application in fractional photonic waveguides and metamaterials described by fractional (nonlocal) Maxwell's equations.\\
\indent
This manuscript is organized as follows. Section \ref{sec_frac} presents the formulation of the space-fractional Schr\"{o}dinger equation (SFSE). Section~\ref{sec_ss} reviews the fundamental concepts of SSs and CPA in non-Hermitian scattering systems. In Section~\ref{sec_nhtt}, we analyze the non-Hermitian scattering dynamics in space-fractional quantum mechanics by examining the dependence of the SS and CPA loci on the L\'{e}vy index $\alpha$, with the corresponding discussion presented therein.
Finally, the paper concludes in Section \ref{sec_con}.

%=============== Section-2 =================
%Rewrite section-2 again: Copy and paste
\section{The fractional Schr\"{o}dinger equation}
\label{sec_frac}
The one-dimensional space-fractional Schr\"{o}dinger equation is given by \cite{laskin2000fractional,laskin20001fractional, laskin2002fractional} 
\begin{equation}
    i\hbar \frac{\partial \psi(x, t)}{\partial t} = \mathcal{H}_\alpha(x, t) \psi(x, t), \qquad 1 < \alpha \le 2,
	\label{frac_01}
\end{equation}
where $\alpha$ denotes the L\'evy index and $\psi(x, t)$ represents the time-dependent wave function. The operator $\mathcal{H}_\alpha(x, t)$ is the fractional Hamiltonian operator defined through the Riesz fractional derivative as
\begin{equation}
	\mathcal{H}_\alpha(x, t) = D_\alpha (-\hbar^2 \Delta)^{\alpha/2} + V(x, t).
    \label{frac_02}
\end{equation}
with $\Delta = \frac{\partial^2}{\partial x^2}$ and the coefficient $D_{\alpha}$ is the generalized fractional quantum diffusion coefficient, which depends on L\'evy index $\alpha$ and having physical dimension $[D_\alpha]=\mathrm{energy}^{1-\alpha}\,\times \mathrm{length}^{\alpha}\,\times \mathrm{time}^{-\alpha}$ \cite{tare2014transmission}.
In the limiting case $\alpha=2$, one recovers the standard Schr\"odinger Hamiltonian with $D_{\alpha} =1/2m$, where $m$ is the particle mass. The Riesz fractional derivative acting on the wave function $\psi(x, t)$ is defined in Fourier space as
\begin{equation}
	(-\hbar^2 \Delta)^{\alpha/2} \psi(x, t) = \frac{1}{2\pi\hbar} \int_{-\infty}^\infty \widetilde{\psi}(p, t)\, |p|^\alpha\, e^{ipx/\hbar}\, dp,
\end{equation}
where $\widetilde{\psi}(p, t)$ is the Fourier transform of $\psi(x, t)$,
\begin{equation}
	\widetilde{\psi}(p, t) = \int_{-\infty}^\infty \psi(x, t)\,e^{-ipx/\hbar}\, dx .
\end{equation}
In the following, we restrict our analysis to time-independent potential, $V(x,t)$ = $V(x)$, for which the fractional Hamiltonian $\mathcal{H}_{\alpha}$ does not depends explicitly on time. Under this condition, separation of variables can be employed by writing the wavefunction as $\psi(x,t)=\psi(x)\,e^{-iEt/\hbar}$. Substituting this form into Eq.~(\ref{frac_01}), one obtains the stationary space-fractional Schrodinger equation
\begin{equation}
	D_\alpha (-\hbar^2 \Delta)^{\alpha/2} \psi(x) + V(x) \psi(x) = E \psi(x),
    \label{frac_stationary}
\end{equation}
where, $E$ denotes the energy eigenvalue. In this study, the generalized fractional diffusion coefficient is taken as 
\begin{equation}
	D_{\alpha}
	= \frac{u^{\,2-\alpha}}{\alpha\,m^{\alpha-1}},
	\label{Dalpha}
\end{equation}
where $u$ represents the characteristic velocity of the non-relativistic quantum system, taken as $1.0\times10^{-5}c$, with $c$ being the speed of light in vacuum (for further details, refer to Section V of Ref.~\cite{dong2007some}).

%================== Section-03 ==================
\section{Spectral singularity and coherent perfect absorption}
\label{sec_ss}
The Hamiltonian operator in one dimension for a non-relativistic particle, expressed in natural units $2m = \hbar = 1$, is given by
\begin{equation}
\mathcal{H} = -\frac{d^2}{dx^{2}} + V(x),
\end{equation}
where the potential is complex, $V(x) = V_{r}(x) + iV_{i}(x)$ with $V_{r}, V_{i} \in \mathbb{R}$, and satisfies the asymptotic condition $V(x) \rightarrow 0$ as $x \rightarrow \pm \infty$. If the function $\mathcal{U}(x) = (1 + |x|)V(x)$ is integrable over the real line, the Hamiltonian admits well-defined scattering solutions with the following asymptotic forms \cite{newton2013scattering,mostafazadeh2009spectral}:
\begin{equation}
\psi(k, x \rightarrow +\infty) = A^{+}(k)e^{ikx} + B^{+}(k)e^{-ikx},
\end{equation}
\begin{equation}
\psi(k, x \rightarrow -\infty) = A^-(k)e^{ikx} + B^-(k)e^{-ikx}.
\end{equation}
where $k = \sqrt{E}$ denotes the wave number. The coefficients $A^{\pm}$ and $B^{\pm}$ are related through the $2 \times 2$ transfer matrix $M(k)$,
\begin{equation}
\begin{bmatrix}
A^{+}(k) \\
B^{+}(k)
\end{bmatrix}
=
M(k)
\begin{bmatrix}
A^{-}(k) \\
B^{-}(k)
\end{bmatrix},
\qquad
M(k)=
\begin{bmatrix}
M_{11}(k) & M_{12}(k) \\
M_{21}(k) & M_{22}(k)
\end{bmatrix}.
\end{equation}
The transmission and reflection amplitudes can be expressed in terms of the transfer-matrix elements as
\begin{equation}
t_l(k) = t_r(k) = \frac{1}{M_{22}(k)}, \qquad r_l(k) = - \frac{M_{21}(k)}{M_{22}(k)}, \qquad r_r(k) = \frac{M_{12}(k)}{M_{22}(k)},
\label{t_and_r}
\end{equation}
SS occurs when $M_{22}(k)$ vanishes at a real wave number $k=k_{s}=\sqrt{E_{s}}$, i.e., 
\begin{equation}
M_{22}(k_{s})=0.
\end{equation}
At this point, the reflection and transmission amplitudes simultaneously diverge, corresponding to a real-energy pole of the scattering coefficients. SSs are therefore associated with zero-width resonances of the eigenvalue equation $\mathcal{H}\psi=k^{2}\psi$
and represent a distinctive feature of non-Hermitian scattering systems \cite{mostafazadeh2009resonance}. The corresponding transmission and reflection coefficients,
\begin{equation}
T_{l/r}=|t_{l/r}|^{2}, \qquad R_{l/r}=|r_{l/r}|^{2},
\end{equation}
also diverge at $E=E_{s}$. Another important phenomenon in NHQM is CPA, which may be considered the time-reversed counterpart of a SS \cite{mostafazadeh2012optical}. CPA occurs at a real positive energy $E=E_{c}$ for which incoming coherent waves are completely absorbed by the system, resulting in vanishing outgoing amplitudes. Within the transfer-matrix formalism, the CPA condition can be expressed as
\begin{equation}
\begin{gathered}
t_{l}(k)\,t_{r}(k) - r_{l}(k)\,r_{r}(k) = 0, \\[6pt]
\Rightarrow \; M_{11}(k_{c}) = 0.
\end{gathered}
\label{cpa_condition}
\end{equation}
where $k_{c} = \sqrt{E_{c}} \in \mathbb{R}^{+}$. At this energy, destructive interference between approximately phased incident waves leads to perfect absorption without reflection or transmission, highlighting the interplay between gain and loss in non-Hermitian systems.

%=============== Section-04 ===============
\section{Non-Hermitian scattering dynamics in space fractional quantum mechanics}
\label{sec_nhtt}
In the previous section, we established the general conditions for the occurrence of SSs and CPA within the standard non-Hermitian scattering framework. These phenomena are characterized by specific properties of the transfer matrix, namely the vanishing of $M_{22}$ and $M_{11}$ terms, respectively. The SFQM generalizes conventional Schr\"odinger dynamics through a fractional kinetic operator and introduces modified scattering properties. Therefore, in this section we investigate the condition for the loci of SSs and CPA within this generalized setting using transfer matrix formalism. We consider a complex potential barrier of the form  $V(x)=V_{r}+iV_{i}$, with $V_{r}, V_{i} \in \mathbb{R}$, defined over the finite interval ($-d, +d$) and zero elsewhere.  The transfer matrix for this potential in SFQM has the following elements \cite{tare2014transmission}
\begin{equation}
\begin{aligned}
M_{11} &= \left[\cos(2\kappa_{\alpha}d) + i\omega_{+}\sin(2\kappa_{\alpha}d)\right]e^{-2ik_{\alpha}d}, \\
M_{22} &= \left[\cos(2\kappa_{\alpha}d) - i\omega_{+}\sin(2\kappa_{\alpha}d)\right]e^{2ik_{\alpha}d},\\
M_{12} &= i \omega_{-} \sin(2\kappa_{\alpha}d), \\
M_{21} &= -i \omega_{-} \sin(2\kappa_{\alpha}d). 
\end{aligned}
\end{equation}
where
\begin{equation}
\omega_{\pm} = \frac{1}{2}\left(\frac{\eta^{2} \pm 1}{\eta} \right), \quad \eta = \left(\frac{k_{\alpha}}{\kappa_{\alpha}}\right)^{\alpha - 1}, \quad k_{\alpha} = \left( \frac{E}{D_{\alpha} \hbar^{\alpha}}\right)^{\frac{1}{\alpha}}, \quad \kappa_{\alpha} = \left( \frac{E-V}{D_{\alpha} \hbar^{\alpha}}\right)^{\frac{1}{\alpha}}.
\label{all_equations_in_one_line}
\end{equation}
The transfer matrix elements satisfy the relations $M_{11}=M_{22}^{*}$ and $M_{12}=M_{21}^{*}$. According to Eq.~(\ref{t_and_r}), a SS occurs at real values of the wave number $k = k_{s}$ for which the transfer-matrix element $M_{22}(k)$ vanishes. 
Imposing the condition $M_{22}(k)=0$, the spectral singularity the SS condition takes the form
%\begin{equation}
%    \cos{2\kappa_{\alpha}d} = i\omega_{+}\sin{2\kappa_{\alpha}d}.
%\end{equation}

%For a complex potential $V$, the above condition reduces to
\begin{equation}    
\cos{2\kappa_{\alpha}d}= \pm \left[ \frac{1+\left(1-\frac{V}{E}\right)^{\frac{2(\alpha-1)}{\alpha}}}{1-\left(1-\frac{V}{E}\right)^{\frac{2(\alpha-1)}{\alpha}}} \right].
\label{cospm}
\end{equation}
Since the potential $V$ is complex, the ratio $V/E$ in Eq.~(\ref{cospm}) is, in general, a complex. For convenience, it may be parameterized as $V/E = \rho+i\sigma$, where $\rho$ and $\sigma$ are real parameters. Consequently, Eq.~(\ref{cospm}) is a complex equation which can be decomposed into a pair of coupled real equations:
\begin{equation}
\begin{aligned}
\cos(\mathbbmss{q})\cosh(\mathbbmss{p}) 
&= \pm \left[ \frac{1 - \lvert \Omega \rvert^{2}}{\lvert 1 - \Omega \rvert^{2}} \right], \\
\sin(\mathbbmss{q})\sinh(\mathbbmss{p}) 
&= \pm \left[ \frac{2\,\mathrm{Im}[\Omega]}{\lvert 1 - \Omega \rvert^{2}} \right].
\end{aligned}
\label{real_imag_eqs}
\end{equation}
Here, the complex quantity $\Omega$ is defined as
\begin{equation}
    \Omega = \left[(1-\rho)^{2}+\sigma^{2} \right]^{\frac{(\alpha-1)}{\alpha}}\, e^{-\frac{2 i (\alpha-1) \theta}{\alpha}}
\end{equation}
and the parameters $\mathbbmss{p}$ and $\mathbbmss{q}$ are given by
\begin{equation}
\mathbbmss{p} =
2k_{\alpha}d\, r^{\frac{1}{\alpha}}
\sin\!\left(\frac{\theta}{\alpha}\right),
\qquad
\mathbbmss{q} =
2k_{\alpha}d\, r^{\frac{1}{\alpha}}
\cos\!\left(\frac{\theta}{\alpha}\right),
\label{peqn_qeqn}
\end{equation}
where 
\[
r=\sqrt{(1-\rho)^2+\sigma^2},
\qquad
\theta=\tan^{-1}\!\left(\frac{\sigma}{1-\rho}\right).
\]
In Eq.~(\ref{real_imag_eqs}), the  same choice of sign ($+$ or $-$) must be adopted consistently in both equation. Solving these equations for $\cos(\mathbbmss{q})$ and $\sin(\mathbbmss{q})$, and utilizing the identity $\sin^{2}(\mathbbmss{q}) + \cos^{2}(\mathbbmss{q}) = 1$, the variable $\mathbbmss{q}$ can be eliminated. Subsequently, using the identity  $\cosh^{2}\mathbbmss{p}=1 + \sinh^{2}(\mathbbmss{p})$, one obtain a quadratic equation in $\sinh^{2}{\mathbbmss{p}}$ of the form
\begin{equation}
\sinh^{4}\mathbbmss{p}+ \left[ 1-  \left( \frac{1-|\Omega|^{2}}{|1-\Omega|^{2}} \right)^{2}-\left( \frac{2\,\mathrm{Im}[\Omega]}{|1-\Omega|^{2}} \right)^{2} \right]\sinh^{2}\mathbbmss{p}-\left( \frac{2\,\mathrm{Im}[\Omega]}{|1-\Omega|^{2}} \right)^{2}=0.
\label{quad}    
\end{equation}
Solving Eq.~(\ref{quad}) for $\mathbbmss{p}$ yields
\begin{equation}
\mathbbmss{p}(\rho,\sigma,\alpha) = \sinh^{-1}\left[\sqrt{\tau(\rho,\sigma,\alpha)}\right]
\label{rquad}
\end{equation}
where $\tau(\rho,\sigma,\alpha)$ corresponds to the physically admissible roots of the quadratic equation. Substituting Eq.~(\ref{rquad}) into the first equation of Eq.~(\ref{real_imag_eqs}) and using the identity $\cosh[\sinh^{-1}(y)]=\sqrt{1+y^{2}}$, leads to a transcendental condition for the parameter $\mathbbmss{q}$. Owing to the periodic nature of the cosine function, this condition admits multiple discrete solutions. For convenience, we introduced the notation
\begin{equation}
\mathbbmss{q}(\rho,\sigma,\alpha,n)
\equiv
\mathbbmss{Q}_{n}^{\pm}(\rho,\sigma,\alpha),
\label{qQ}
\end{equation}
where $n\in\mathbb{Z}$, and 
\begin{equation}
\mathbbmss{Q}_{n}^{\pm}(\rho,\sigma,\alpha)= \pi n  \pm  \cos^{-1}\!\left[
\frac{1-|\Omega|^{2}}{|1-\Omega|^{2}}
\frac{1}{\sqrt{1+\tau(\rho,\sigma,\alpha)}}
\right].
\label{Qeq}
\end{equation}
Using Eqs.~(\ref{peqn_qeqn}) and (\ref{qQ}), one obtains
\begin{equation}
\frac{\mathbbmss{P}_{n}^{\pm}(\rho,\sigma,\alpha)}
{\mathbbmss{Q}_{n}^{\pm}(\rho,\sigma,\alpha)}
=
\tan\!\left(\frac{\theta}{\alpha}\right),
\label{Peq}
\end{equation}
where $\mathbbmss{P}_{n}^{\pm}(\rho,\sigma,\alpha) \equiv \mathbbmss{p}(\rho,\sigma,\alpha, n)$. 
Eq.~(\ref{Peq}) generates an infinite family of conditions defining the loci of SSs in $\rho$-$\sigma$ plane for a fixed value of L\'{e}vy fractional parameter $\alpha$, with each integer $n$ corresponding to a distinct branch of solutions. These, loci may equivalently be characterized by introducing the function
\begin{equation}
\mathbbmss{S}_{n}^{\pm}(\rho,\sigma,\alpha)
=
\mathbbmss{p}(\rho,\sigma,\alpha)
-
\mathbbmss{P}_{n}^{\pm}(\rho,\sigma,\alpha).
\end{equation}
Such that the SS condition is given by $\mathbbmss{S}_{n}^{\pm}(\rho,\sigma,\alpha)=0$. This representation provides a compact way to identify the admissible parameter sets associated with SSs. To determine the allowed values of the integer index $n$ and the appropriate branch $(\pm)$, we express the quantity $k_{\alpha}$ in terms of the previously defined parameters. From Eq.~(\ref{peqn_qeqn}), we obtains  
\begin{equation}
k_{\alpha}d
=
\frac{\mathbbmss{q}}
{2\,r^{1/\alpha}\cos(\theta/\alpha)}
=
\frac{\mathbbmss{Q}_{n}^{\pm}(\rho,\sigma,\alpha)}
{2\,r^{1/\alpha}\cos(\theta/\alpha)}
\equiv
\mathbbmss{H}_{n}^{\pm}(\rho,\sigma,\alpha),
\label{hn}
\end{equation}
which explicitly relates the SS condition to the underlying system parameters Substituting Eq.~(\ref{hn}) into Eq.~(\ref{cospm}) yields a family of complex equation describing the curves in the $\rho$--$\sigma$ plane along which SSs occur. These curves may be compactly as
$\widetilde{\mathbbmss{S}}_{n}^{\pm}(\rho,\sigma,\alpha)=0$, where
\begin{align}
\widetilde{\mathbbmss{S}}_{n}^{\pm}(\rho,\sigma,\alpha)
&=
e^{-2i \mathbbmss{H}_{n}^{\pm}(\rho,\sigma,\alpha)\left\{(1-\rho)-i \sigma\right\}^{1/\alpha}}
\left[1+\left\{(1-\rho)-i \sigma\right\}^{(\alpha-1)/\alpha} \right]^{2}
\nonumber \\
&\quad
-
e^{2i \mathbbmss{H}_{n}^{\pm}(\rho,\sigma,\alpha)\left\{(1-\rho)-i \sigma\right\}^{1/\alpha}}
\left[1-\left\{(1-\rho)-i \sigma\right\}^{(\alpha-1)/\alpha} \right]^{2}.
\label{Sn_tilde}
\end{align}
It is found that $\lvert \widetilde{\mathbbmss{S}}_{n}^{\pm}(\rho,\sigma,\alpha) \rvert = 0$ only for $n>0$ when the negative branch ($-$) is selected. This result generalizes the SS condition obtained in the standard quantum-mechanical limit.
\begin{figure}[t]
    \centering
    \includegraphics[width=0.95\linewidth]{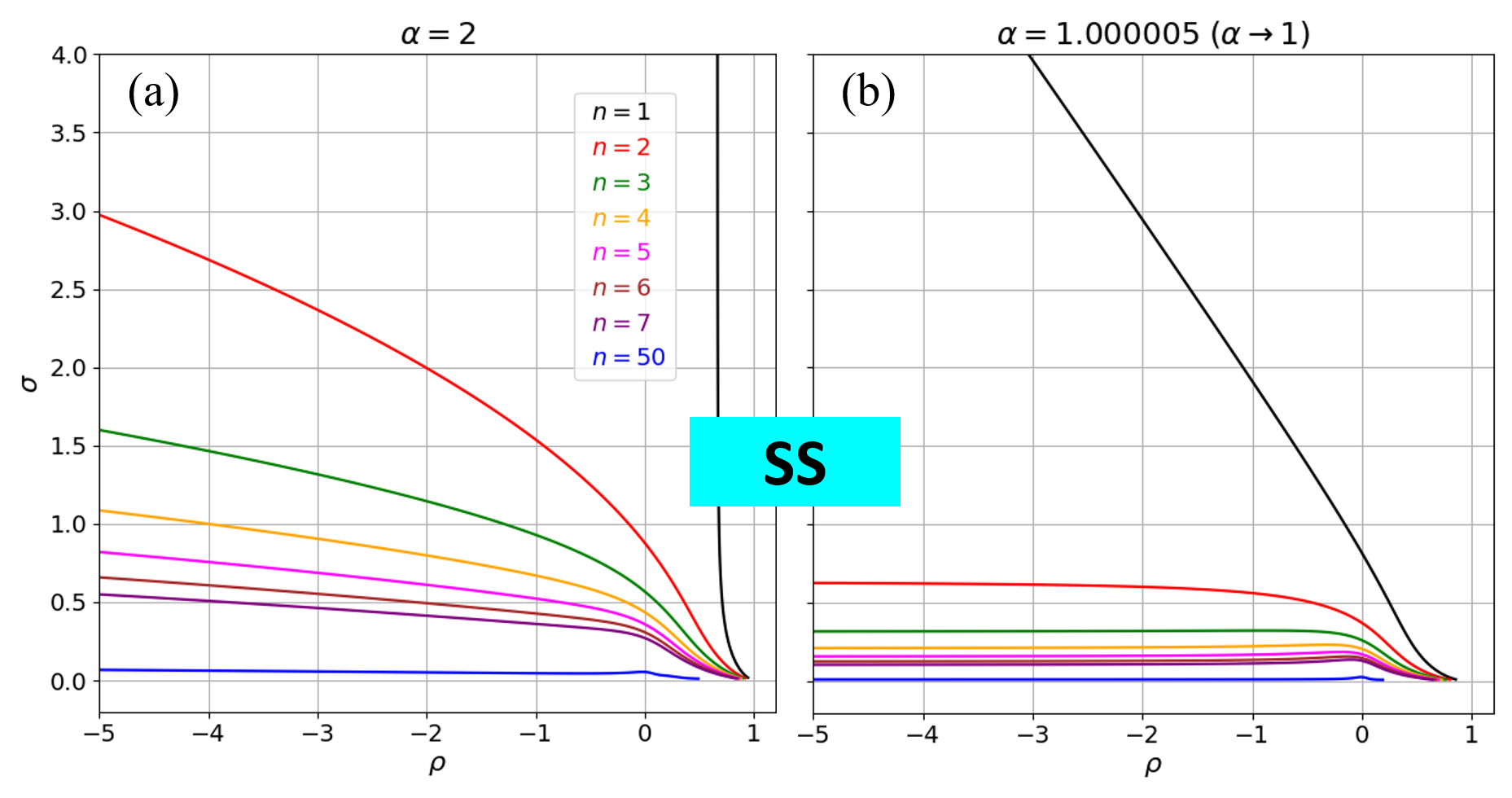}
    \caption{(Color online). Graph of the curve $\widetilde{\mathbbmss{S}}_{n}^{-}(\rho,\sigma,\alpha)$ in the $\rho-\sigma$ plane for different $n$ with the L\'{e}vy index (a) $\alpha=2$ and (b) $\alpha=1.000005$ ($\alpha \rightarrow 1$). The result shown in (a) is fully consistent with the findings reported in Ref. \cite{mostafazadeh2009resonance}. It is evident from the plots that no spectral singularity exists for $\sigma < 0$, as expected. Although for both cases, all the curves merge at $\rho=1$ for $\sigma=0$, this point does not correspond to a spectral singularity, since the potential becomes real when $\sigma=0$. $\widetilde{\mathbbmss{S}}_{1}^{-}(\rho,\sigma,\alpha)$ has a vertical asymptote at $\rho=0.667$ in figure (a).}
    \label{ss_01}
\end{figure}
For $\alpha=2$, the Eq. (\ref{Sn_tilde}) reduces to
\begin{align} \nonumber
	 \widetilde{\mathbbmss{S}}_{n}^{\pm}(\rho,\sigma,2) = e^{-2i \mathbbmss{H}_{n}^{\pm}(\rho,\sigma,2)\sqrt{(1-\rho)-i \sigma}}\left[1+\sqrt{(1-\rho)-i \sigma}\right]  \\  -  e^{2i \mathbbmss{H}_{n}^{\pm}(\rho,\sigma, 2)\sqrt{(1-\rho)-i \sigma}}\left[1-\sqrt{(1-\rho)-i \sigma}\right] 	
     \label{Sn_tilde2}
\end{align}
where $\mathbbmss{H}_{n}^{\pm}(\rho,\sigma,2)$ can be determined from Eq. (\ref{hn}). and this is fully consistent with that reported in \cite{mostafazadeh2009resonance}. Fig. \ref{ss_01}a depicts the family of curves in the $\rho-\sigma$ plane for the L\'{e}vy index $\alpha=2$ (standard QM), defined by the condition $\widetilde{\mathbbmss{S}}_{n}^{-}(\rho,\sigma,\alpha)=0$, for different values of $n$.\\
\indent
The loci of SSs lie entirely in the region $\sigma>0$, exhibiting the important fact that spectral singularities can occur only for a positive coupling constant. The function $\widetilde{\mathbbmss{S}}_{1}^{-}(\rho,\sigma,\alpha)$ exhibits a vertical asymptote at $\rho = 0.667$. It is also evident that, as $n$ increases ($n \rightarrow \infty$), the curves progressively flatten and approach the $\rho$-axis, indicating that the magnitude of $\sigma$ associated with SSs diminishes in the large-$n$ limit. This behavior can also be shown analytically from Eq.~(\ref{Sn_tilde}). All curves intersect at the common point $(\rho,\sigma)=(1,0)$, reflecting a universal limiting behavior independent of $n$. However, this point does not correspond to an SS, since the potential becomes purely real when $\sigma=0$. For all $n\geq1$, the $\widetilde{\mathbbmss{S}}_{n}^{-}(\rho,\sigma,\alpha)$ exhibiting the monotonically decresing behaviour when $\rho$ increases. Also, for $n\geq2$, the graph of $\widetilde{\mathbbmss{S}}_{n}^{-}(\rho,\sigma,\alpha)$ lies above the graph of $\widetilde{\mathbbmss{S}}_{n+1}^{-}(\rho,\sigma,\alpha)$.\\
\begin{figure}[!t]
    \centering
    \includegraphics[width=1.07\linewidth]{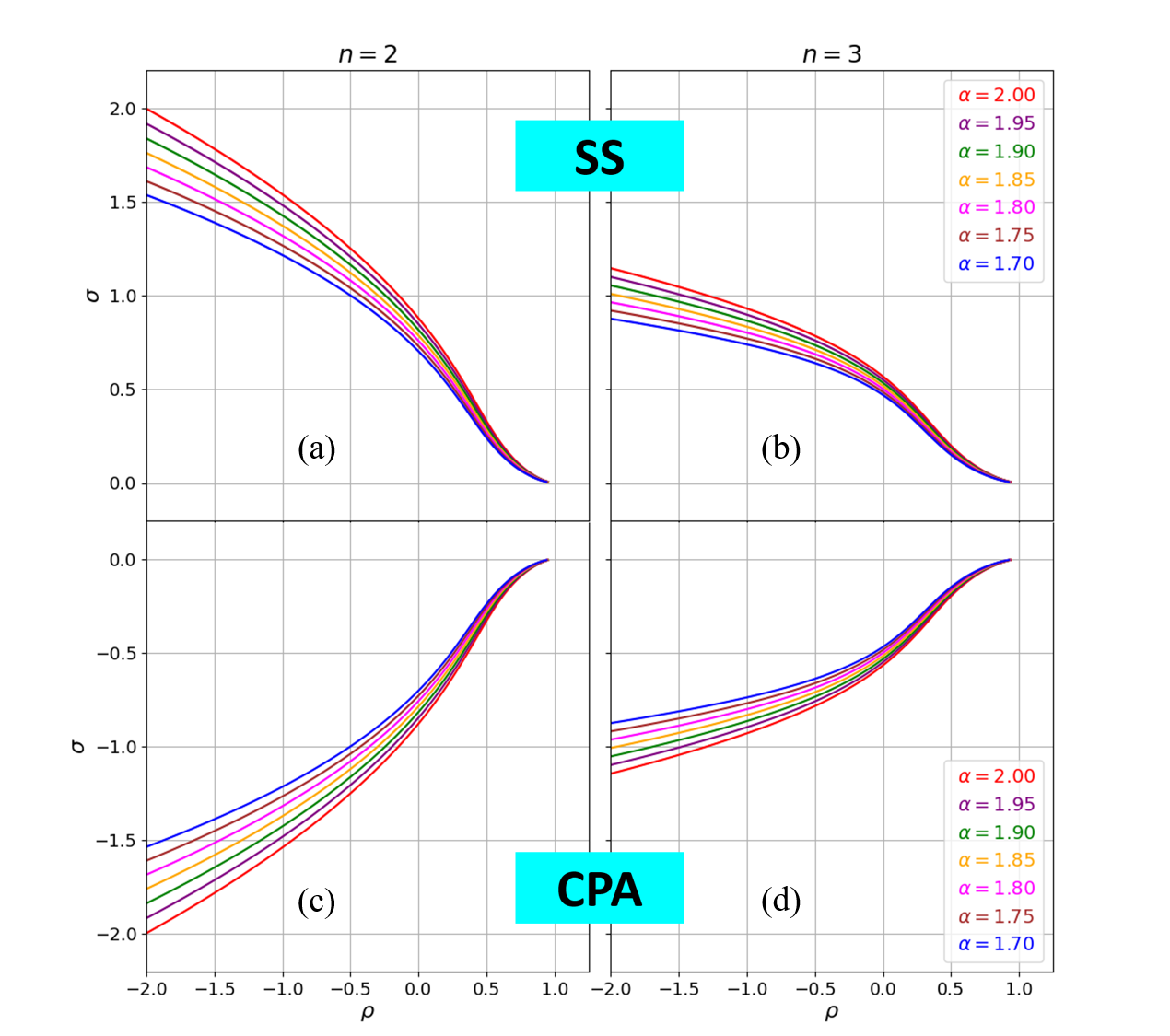}
    \caption{(Color online). Illustration of the loci of the SSs defined by $\widetilde{\mathbbmss{S}}_{n}^{-}(\rho,\sigma,\alpha)$ in the $\rho-\sigma$ plane for (a) $n=2$ and (b) $n=3$, shown for different values of the L\'{e}vy index $\alpha$. Panels (c) and (d) display the corresponding loci of CPA, the time-reversed counterpart of SS, defined by $\widetilde{\mathbbmss{C}}_{n}^{-}(\rho,\sigma,\alpha)$ for $n=2$ and $n=3$, respectively, for the same set of $\alpha$ values. As expected from time-reversal symmetry, the SS loci are confined to the region $\sigma>0$, while the CPA loci occur for $\sigma<0$, indicating that SSs and CPA solutions arise for opposite signs of the non-Hermitian coupling.}
    \label{ss_cpa}
\end{figure}
\indent
Fig. \ref{ss_01}b illustrates the corresponding SS loci for the same set of mode indices $n$, but for a L\'{e}vy index very close to unity, $\alpha = 1.000005$. Compared to the standard QM case ($\alpha=2$), a marked qualitative and quantitative modification of the SS loci structure is observed. In particular, for all values of $n>1$, all SS curves are strongly compressed and flattened toward the $\rho$-axis, showing a substantial reduction in the magnitude of $\sigma$ required for the occurrence of spectral singularities. This feature implies that the dependence of the SS condition on $\rho$ is substantially weakened in this regime. Despite these quantitative changes, the qualitative trends observed for $\alpha=2$ remain preserved. The SS loci are confined entirely to the region $\sigma>0$, and increasing $n$ continues to produce progressively flatter curves that approach the $\rho$-axis in the large-$n$ limit. The curves follow the same trend of converging to the point $(\rho,\sigma)=(1,0)$ as in Fig.~\ref{ss_01}a. However, the approach to this point is much more gradual in the fractional case. In this case, for all $n \geq 1$, the graph of $\widetilde{\mathbbmss{S}}_{n}^{-}(\rho,\sigma,\alpha)$ lies above the $\widetilde{\mathbbmss{S}}_{n+1}^{-}(\rho,\sigma,\alpha)$.\\
\indent
Fig. \ref{ss_cpa}a and \ref{ss_cpa}b illustrate the dependence of the SS loci on the L\'{e}vy index $\alpha$ 
for fixed values of $n=2$ and $n=3$, respectively. Within the framework of fractional 
quantum mechanics, the parameter $\alpha$ governs the degree of nonlocality of the 
fractional kinetic operator and interpolates between strongly nonlocal dynamics for 
$\alpha<2$ and the standard Schr\"{o}dinger limit at $\alpha=2$. For a fixed $n$, 
decreasing $\alpha$ systematically shifts the SS curves toward smaller values of 
$\sigma$ across the entire range of $\rho$. This behavior shows that an enhancement 
of the fractional character of the dynamics reduces the magnitude of the non-Hermitian 
coupling required for the emergence of SSs. Physically, stronger 
L\'{e}vy-flight effects associated with smaller $\alpha$ enhance long-range transport and 
interference, thereby promoting the threshold scattering behavior characteristic of 
SSs and allowing SSs to occur at weaker gain-loss strengths. This 
trend is consistently observed for both $n=2$ and $n=3$, and it becomes pronounced 
with increasing $n$. In particular, for the same value of $\alpha$, the SS loci 
corresponding to $n=3$ lie below those for $n=2$, in agreement with Fig. \ref{ss_01}, 
where increasing $n$ leads to a progressive flattening of the curves and a suppression 
of $\sigma$ in the large-$n$ limit.\\
\indent
Taken together with Fig. \ref{ss_01}, these results demonstrate an interplay 
between the L\'{e}vy index $\alpha$ and the parameter $n$. Increasing nonlocality 
(smaller $\alpha$) lowers the critical non-Hermiticity required for the formation of 
SSs. A similar reduction in the required gain-loss strength is observed upon increasing 
the parameter $n$, which leads to a progressive flattening of the SS loci and a 
suppression of the corresponding $\sigma$ values. The combined effect of a larger $n$ 
and a smaller $\alpha$ therefore acts cooperatively to facilitate the emergence of 
SSs at weaker non-Hermitian coupling. Despite these modifications 
to the SS landscape, all SSs loci remain strictly confined to the 
region $\sigma>0$, in agreement with the requirement of positive coupling strength for 
SS formation. Furthermore, for all values of $n$ and $\alpha$, the SS loci converge 
to the universal point $(\rho,\sigma)=(1,0)$, reflecting a limiting behavior independent 
of these parameters. However, this point does not correspond to a SS, since the potential 
becomes purely real when $\sigma=0$, and the non-Hermitian mechanism responsible for SS 
formation is absent. For a fixed value of $n$, the curve 
$\widetilde{\mathbbmss{S}}_{n}^{-}(\rho,\sigma,\alpha_1)$ always lies above 
$\widetilde{\mathbbmss{S}}_{n}^{-}(\rho,\sigma,\alpha_2)$ for 
$\alpha_1 > \alpha_2$.
 \\
\indent
An important observation reported in \cite{hasan2018new} is that the energy $E_{S}(=E_{ss})$ at which a SS occurs for a complex potential $V_r + iV_i$, with fixed values of $V_r$ and $V_i$, undergoes a blue shift as the fractional parameter $\alpha$ decreases. Our results for the SS loci as a function of $\alpha$ are fully consistent with this finding. In this work, the SS loci are analyzed in terms of the dimensionless parameters $\rho = V_r/E$ and $\sigma = V_i/E$. For fixed potential strengths $V_r$ and $V_i$, a decrease in the values of $\rho$ and $\sigma$ necessarily implies an increase in the corresponding energy $E$. As shown in Figs. \ref{ss_cpa}a and \ref{ss_cpa}b, decreasing the fractional index $\alpha$ systematically shifts the SS loci toward smaller values of $\rho$ and $\sigma$. Since the potential parameters are held fixed, this shift can only be achieved by increasing the energy associated with the spectral singularity. The energy $E$ appearing in the definitions of $\rho$ and $\sigma$ is the energy at which the SS occurs, i.e., $E = E_{S}$. The suppression of $\rho$ and $\sigma$ with decreasing $\alpha$ implies an increase in $E_{ss}$. Consequently, the SS energy experiences a blue shift as the fractional parameter $\alpha$ decreases, in direct agreement with the results reported in \cite{hasan2018new}. This provides an independent and geometrically transparent confirmation of the fractional-energy blue shift of SSs within the locus framework presented in this work.\\
\indent
Next, in order to calculate the locus condition for CPA we put $M_{11}=0$ and applying the similar methodology to derive the expression for $\widetilde{\mathbbmss{C}}_{n}^{\pm}(\rho,\sigma,\alpha)=0$. This expression gives an infinite sequence of curves in the $\rho-\sigma$ plane, where the CPAs resides and is expressed as 
\begin{align} \nonumber
	 \widetilde{\mathbbmss{C}}_{n}^{\pm}(\rho,\sigma,\alpha) = e^{2i \mathbbmss{H}_{n}^{\pm}(\rho,\sigma,\alpha)\left\{(1-\rho)-i \sigma\right\}^{\frac{1}{\alpha}}}\left[1+\left\{(1-\rho)-i \sigma\right\}^{\frac{(\alpha-1)}{\alpha}} \right]^{2} \\ -  e^{-2i \mathbbmss{H}_{n}^{\pm}(\rho,\sigma,\alpha)\left\{(1-\rho)-i \sigma\right\}^{\frac{1}{\alpha}}}\left[1-\left\{(1-\rho)-i \sigma\right\}^{\frac{(\alpha-1)}{\alpha}} \right]^{2} 	
     \label{Cn_tilde}
\end{align}
Following a similar approach as demonstrated in Eq. (\ref{hn}), we define $\mathbbmss{H}_{n}^{\pm}(\rho,\sigma,\alpha)$ in this particular context. Analogous to SSs, the CPA condition $|\widetilde{\mathbbmss{C}}_{n}^{\pm}(\rho,\sigma,\alpha)| = 0$
is realized only for $n>0$ and taking minus ($-$) only. Fig. \ref{ss_cpa}c and \ref{ss_cpa}d depict the loci of CPA, described by $\widetilde{\mathbbmss{C}}_{n}^{-}(\rho,\sigma,\alpha)$, in the $\rho-\sigma$ plane for $n=2$ and $n=3$, respectively, for different values of the L\'{e}vy index $\alpha$. Since CPA represents the time-reversed counterpart of SSs, the corresponding solutions emerge only in the presence of loss, and consequently all CPA curves are confined to the region $\sigma<0$. For a fixed value of $n$, decreasing $\alpha$ shifts the CPA curves toward larger negative values of $\sigma$ over the entire range of $\rho$. This behavior mirrors the SS case, where decreasing $\alpha$ lowers the positive $\sigma$ required for SS formation. In the CPA case, enhanced nonlocality associated with smaller $\alpha$ strengthens long-range transport and interference, thereby allowing perfectly absorbing states to occur at weaker loss strengths. The monotonic increase of $\sigma$ with $\rho$ observed in Fig. \ref{ss_cpa}c and \ref{ss_cpa}d is the exact sign-reversed counterpart of the monotonic decrease observed for SSs in Fig. \ref{ss_cpa}a and \ref{ss_cpa}b, respectively.\\
\indent
A comparison between Fig. \ref{ss_cpa}c and \ref{ss_cpa}d further reveals that, for the same value of $\alpha$, the CPA loci corresponding to $n=3$ lie closer to the $\rho$-axis than those for $n=2$. This trend is consistent with the behavior observed for SSs, where increasing $n$ leads to a flattening of the curves and a suppression of the magnitude of $\sigma$. Thus, increasing the mode index $n$ reduces the magnitude of loss required to achieve CPA. Taken together with the SS panels, these results demonstrate the SS-CPA duality under time reversal: while SSs occur for $\sigma>0$ and correspond to purely outgoing threshold states, CPA arises for $\sigma<0$ and corresponds to purely incoming, perfectly absorbed states. The symmetric structure of the SS and CPA loci with respect to the $\sigma=0$ axis demonstrates the underlying reversal relationship between them.
\section{Conclusion}
\label{sec_con}
In this study, we investigated how SSs and their time-reversed counterpart, CPA, emerge and evolve in non-Hermitian space fractional quantum mechanics. Starting from a complex scattering potential, we derived the conditions for SS and CPA and represented them geometrically as curves in a dimensionless parameter space defined by the ratios of the real and imaginary parts of the potential to the energy. This representation allowed us to directly visualize how fractional dynamics, encoded through the L\'{e}vy index $\alpha$, modify the non-Hermitian scattering behavior. We showed that reducing $\alpha$, which enhances nonlocal transport, systematically lowers the gain-loss strength required for the formation of SSs and CPA, while increasing the mode index $n$ further suppresses this threshold. Importantly, we demonstrated that decreasing the fractional parameter $\alpha$ induces a blue shift of the SS energy for fixed potential strengths, providing a clear and intuitive explanation consistent with earlier studies. Beyond its quantum-mechanical setting, the present study may find application in fractional photonic waveguides and metamaterials governed by fractional Maxwell's equations. In such systems, the propagation constant $\beta$ plays the role of the spectral parameter, while gain and loss are introduced through the complex refractive index $\Delta n_r + i\Delta n_i$, leading to the dimensionless ratios $\Delta n_r/\beta$ and $\Delta n_i/\beta$, which directly translate the parameters $\rho = V_r/E$ and $\sigma = V_i/E$ defined in this work. From this viewpoint, our results offer qualitative insight into the control of lasing thresholds and coherent perfect absorption in fractional waveguides and identify the index $\alpha$ as a tunable nonlocal control knob for SS-CPA physics in both non-Hermitian quantum systems and wave-based fractional systems.\\
\\
\\
\\
\\
\\
\noindent
{\it \bf{Acknowledgments}}:\\
\\
VNS acknowledges the support from the Department of Physics, School of Sciences, IILM University, Greater Noida, for providing a conducive research environment. MU sincerely thanks Prof. P. Senthilkumaran, Optics and Photonics Centre (OPC), Indian Institute of Technology Delhi (IIT Delhi), for his constant encouragement for research activities. 

%==================== References =========================
\newpage
\bibliographystyle{unsrturl}
\bibliography{References}

\end{document}